\documentclass[ALICE,manyauthors]{cernphprep}

\newcommand{\jpsi}{\rm J/$\psi$}

\usepackage{rotating}
\usepackage{hyperref}

\begin{document}%
%
%
\begin{titlepage}
\PHnumber{2011-057}      
\PHdate{23 October 2012}              
%
%
\title{Rapidity and transverse momentum dependence  
of inclusive 
J/$\mathbf{\psi}$ production
in pp collisions at $\mathbf{\sqrt{s}}$=7 TeV}
\ShortTitle{Rapidity and $p_{\rm T}$ dependence of inclusive 
J/$\mathbf{\psi}$ production at 7 TeV}   
%
\Collaboration{ALICE Collaboration%
         \thanks{See Appendix~\ref{app:collab} for the list of collaboration 
                      members}}
\ShortAuthor{ALICE Collaboration}      
\begin{abstract}
  The ALICE experiment at the LHC has studied inclusive \jpsi\ production at central
  and forward rapidities in pp collisions at $\sqrt{s}$~=~7~TeV. 
  In this Letter, we report on the first results obtained detecting
  the \jpsi\ through the dilepton decay into e$^+$e$^-$ and $\mu^+\mu^-$
  pairs in the rapidity ranges $|y|<0.9$ and $2.5<y<4$, respectively,  
  and with acceptance down to zero $p_{\rm T}$.
  In the dielectron channel the analysis was carried out on a data
  sample corresponding to an integrated luminosity $L_{\rm int}$~=~5.6~nb$^{-1}$
  and the number of signal events is $N_{\rm J/\psi}$~=~352~$\pm$~32~(stat.)~$\pm$~28~(syst.);
  the corresponding figures in the dimuon channel are $L_{\rm int}$~=~15.6~nb$^{-1}$ and  
  $N_{\rm J/\psi}$~=~1924~$\pm$~77~(stat.)~$\pm$~144~(syst.).
  The measured production cross sections are $\sigma_{\rm
    J/\psi}(|y|<0.9)$~=~ 
    12.4~$\pm$~1.1~(stat.)~$\pm$~1.8~(syst.)~$^{+ 1.8}_{-2.7}$~(syst.pol.) $\mu$b and
  $\sigma_{\rm J/\psi}(2.5<y<4)$~=~6.31~$\pm$~
  0.25~(stat.)~$\pm$~0.76~(syst.)~$^{+0.95}_{-1.96}~$(syst.pol.)  $\mu$b. 
  The differential cross sections, in transverse momentum and
  rapidity, of the \jpsi\ were also measured.
\end{abstract}
\end{titlepage}
\setcounter{page}{2}
\section{Introduction}

The hadroproduction of heavy quarkonium states is governed by both perturbative and non-perturbative aspects of Quantum Chromodynamics (QCD) and was extensively studied at the Tevatron~\cite{Aco05,Abu07,Aba96,Abb99} and RHIC~\cite{Ada07} hadron colliders. Various theoretical approaches, recently reviewed in~\cite{QWG05,Lan09}, were proposed to describe the data. They mainly differ in the details of the non-perturbative evolution of the heavy quark pair towards a bound state. The models are not able to consistently reproduce the production cross section, the transverse momentum ($p_{\rm T}$) distributions and the polarization. Recently, theoretical studies focused on 
the calculation of NLO and NNLO contributions, finding
that their impact on the results is quantitatively
important~\cite{Cam07,Gon08,Gon09,Bro09,But11}. Measurements in the new energy 
domain of the LHC are crucial for a deeper
understanding of the physics involved in hadroproduction
processes. Furthermore, the range of Bjorken-$x$ values accessible at LHC 
energies is unique. Low-$p_{\rm T}$ charmonium measurements, 
in particular at forward rapidity, are sensitive to an unexplored  
region ($x<$10$^{-5}$ at $Q^2=m^2_{\rm J/\psi}$) of the gluon distribution 
function of the proton.

Heavy quarkonia are measured in the ALICE experiment \cite{Aam08} 
through their e$^+$e$^-$ and $\mu^+\mu^-$ decays.  In this Letter we
present the results on inclusive \jpsi\ production in pp collisions at
$\sqrt{s}$~=~7~TeV, measured in the rapidity regions $|y|<0.9$ for
the dielectron channel and
$2.5<y<4.0$ for the dimuon decay channel\footnote{The muon spectrometer covers,
in the ALICE official reference frame, a negative $\eta$ range and, 
consequently, a negative $y$ range. 
However, since in pp the physics is symmetric with respect to $y$=0, we have 
dropped the negative sign when quoting rapidity values.}.
First, a description of the ALICE experimental apparatus is given, 
mainly focusing on the muon detection. Track reconstruction in the central
rapidity region was discussed previously~\cite{alice_pt}.
Details are provided concerning the data analysis, the reconstruction algorithm,
the event selection criteria and the techniques used for the
extraction of the signal. After describing the
determination of the acceptance and efficiency corrections, and the
methods used for the evaluation of the luminosity, the 
values of the integrated, $y$-differential and $p_{\rm T}$-differential 
\jpsi\ cross sections are presented and compared with the results obtained by 
the other LHC experiments~\cite{Kha10,Aai11,Aad11}.

\section{Experimental apparatus and data taking conditions}

The ALICE experiment~\cite{Aam08} consists of two main parts: a central barrel 
and a muon spectrometer. The central barrel detectors ($|\eta |<$0.9) are  
embedded in a large solenoidal magnet, providing a magnetic field of 0.5~T.
Various detector systems track particles down to $p_{\rm T}$ of about 
100 MeV/$c$, and can provide particle identification over a wide momentum range. 
The muon spectrometer covers the pseudorapidity range
$-4<\eta<-2.5$ and
detects muons having $p>$4 GeV/$c$. Finally, various sets of forward
detectors further extend the charged particle pseudorapidity coverage 
up to $\eta = 5.1$ and can 
be used for triggering purposes.

The barrel detectors used in this analysis are the
Inner Tracking System (ITS) and the Time Projection Chamber (TPC).
The ITS \cite{Aam08,alice_its} is a cylindrically-shaped silicon tracker that 
surrounds the central beam pipe. It consists of six layers, with radii between 
3.9 cm and 43.0 cm, covering the pseudo-rapidity range $|\eta|<0.9$. The
two innermost layers are equipped with Silicon Pixel Detectors (SPD),
the two intermediate layers contain Silicon Drift Detectors (SDD),
and Silicon Strip Detectors (SSD) are used on the two outermost
layers. The main task of the ITS is to provide precise track and
vertex reconstruction close to the interaction point, to improve
the overall momentum resolution and to extend tracking down to very low 
$p_{\rm T}$. The SPD can also deliver a signal 
for the first level trigger (L0), which is based
on a hit pattern recognition system at the level of individual
readout chips. 

The TPC~\cite{alice_tpc} is a large cylindrical drift detector
with a central high voltage membrane maintained at $-100$~kV and two readout 
planes at the end-caps. The active volume extends over the ranges $85<r<247$~cm and $-250 < z < 250$~cm in the radial and longitudinal (beam) directions, 
respectively.
Besides being the main tracking detector in the central barrel, the TPC also provides charged hadron identification with very good purity up to a total momentum of about 
3 GeV/c~\cite{alice-pid900} and electrons up to about 10 GeV/c, via specific energy loss (d$E$/d$x$) measurement, with a resolution of $\sigma$=5.5\% for minimum ionizing particles.

Other central barrel detectors with full azimuthal coverage, in particular the 
Time-Of-Flight (TOF)~\cite{Aki10}, the Transition Radiation Detector 
(TRD)~\cite{Kwe09} and the Electromagnetic Calorimeter (EMCAL)~\cite{Abe10}, 
are not used in this analysis, but are expected to significantly improve the 
electron identification and triggering capabilities of the experiment in the 
future.

The muon spectrometer consists of a front absorber followed
by a 3 T$\cdot$m dipole magnet, coupled to tracking and triggering
devices. Muons emitted in the forward rapidity region are filtered 
by means of a 10 interaction length ($\lambda_{\rm I}$) thick front absorber 
made of carbon, concrete and steel, and placed between 0.9 and 5.0 m 
from the nominal position of the interaction point (IP).
Muon tracking is performed by means of 5 tracking stations, positioned
between 5.2 and 14.4 m from the IP, each one based on two planes of
Cathode Pad Chambers. The total number of electronic channels is close
to $1.1\cdot 10^6$, and the intrinsic spatial resolution for these 
detectors is of the order of 70 $\mu$m in the bending direction. 
Stations 1 and 2 (4 and 5) are located upstream (downstream) of 
the dipole magnet,
while station 3 is embedded inside its gap.  A muon triggering system
is placed downstream of a 1.2 m thick iron wall (7.2 $\lambda_{\rm I}$), which 
absorbs secondary hadrons escaping the front absorber and low-momentum
muons (having $p<1.5$ GeV/$c$ at the exit of the front absorber). 
It consists of two stations positioned at 16.1 and 17.1 m from
the IP, equipped with two planes of Resistive Plate Chambers (RPC)
each. The spatial resolution achieved is better than 1 cm, while the
time resolution is of the order of 2 ns.  Throughout its full length,
a conical absorber ($\theta< 2^{\circ}$) made of tungsten, lead and steel
protects the muon spectrometer against secondary particles produced
by the interaction of large-$\eta$ primaries in the beam pipe.

Finally, the VZERO detector consists of two scintillator arrays
covering the range $2.8<\eta < 5.1 $ and $-3.7 <\eta <-1.7$, and positioned,
respectively, at $z=340$ and $z=-90$ cm from the IP. It provides timing 
information to the L0 trigger with a resolution better than 1 ns. This feature 
proves to be useful in the offline rejection of beam-halo and beam-gas events.

The results presented in this Letter were obtained by analyzing  
data collected in the first year of operation of the
LHC, corresponding to pp collisions at $\sqrt{s}$=7 TeV. During this period, 
the LHC reached its goal of delivering more than 10$^{32}$ 
cm$^{-2}s^{-1}$  instantaneous luminosity. 
In ALICE, the instantaneous luminosity was kept to  
$0.6-1.2\cdot 10^{29}$ cm$^{-2}$ s$^{-1}$  in order to have a 
collision pile-up rate in the same bunch crossing below 5\%.
In order to increase the statistics for low cross-section processes,
ALICE ran, during short periods, at 
luminosities about 10 times higher, therefore with a much larger 
pile-up rate in the same bunch crossing. 
For the analysis presented in this Letter, the collected data
were divided into three sub-periods. Each one corresponds 
to similar average luminosity and pile-up rates,
and is characterized by reasonably stable tracking and trigger detector 
configuration and performance. 

The event sample used in this analysis corresponds to minimum bias events
(MB trigger) and, for the muon analysis, to events where 
the detection of at least one muon in the angular acceptance of the muon 
spectrometer ($\mu$-MB trigger) is additionally required.  The MB trigger is
defined as the logical OR between the requirement of at least one fired
readout chip in the SPD, and a signal in at least one of the two VZERO
detectors~\cite{Aam10}. It also requires a coincidence with signals
from two beam pick-up counters, one on each side of the interaction
region, indicating the passage of proton bunches.  The $\mu$-MB
trigger allows the selection of events where at least one particle was 
detected in the trigger chambers of the muon spectrometer. The
trigger logic is based on the requirement of having at least 3 hits
(out of 4) in the trigger stations, both in the bending and
non-bending directions~\cite{Gue06}. In this way one can define a
``trigger track'', and compute its deviation with respect to a track
with infinite momentum.  By requiring such a deviation to be smaller than
a certain value one can select muon candidate tracks having a
transverse momentum larger than a pre-defined value. Such a 
$p_{\rm T}^{\rm {trig}}$ cut can be used to reject soft muons, dominated by 
$\pi$ and K decays, and is able to limit the muon trigger rate when the
machine luminosity is high. The instantaneous luminosity at ALICE
allowed for the choice of the lowest $p_{\rm T}^{\rm {trig}}$ threshold 
(0.5 GeV/$c$), leading to a $\mu$-MB trigger rate between 30 and 500 Hz.
With this $p_{\rm T}^{\rm {trig}}$, the effect of the  
trigger response function on the \jpsi\ detection efficiency is negligible.  
Note that the
effect of such a cut is not sharp and that the selection efficiency
reaches the plateau value only at $p_{\rm T}\sim$1.5 GeV/$c$.
Finally, in order to limit the systematic uncertainties related to 
non-uniformities in the detector response, data were collected by periodically 
varying the polarities of the solenoidal and dipole magnets.

\section{Data analysis}

For the dielectron analysis, 3.5$\cdot$10$^8$ minimum bias 
events ($N_{\rm {MB}}$) are analyzed.  An event with a reconstructed
vertex position $z_{\rm v}$ is accepted if $|z_{\rm v}|<10$~cm.
The tracks are required to have a minimum $p_{\rm T}$ of 1.0~GeV/$c$, a minimum
number of 70 TPC clusters per track (out of a maximum of 159), a $\chi^2$ per 
space point of the momentum fit lower than 4, and to point back to the 
interaction vertex within 1~cm in the transverse plane.  
A hit in at least one of the
two innermost layers of the ITS is required in order to reduce the contribution 
of electrons from $\gamma$ conversions.  For full-length tracks, the 
geometrical coverage of the central barrel detectors is $|\eta|<0.9$.

\begin{figure}[htbp]
\centering
{\includegraphics[width=.6\textwidth]{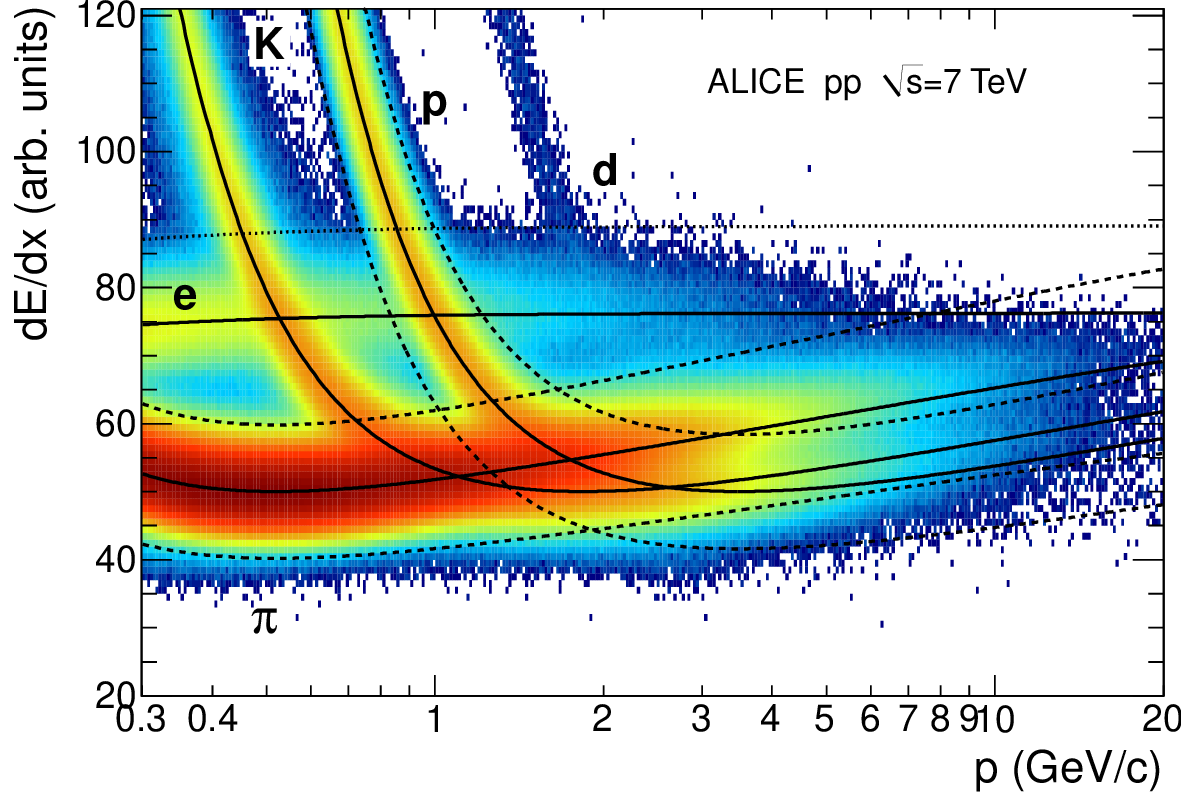}}
\caption{Specific energy loss in the TPC as a function of momentum with 
superimposed Bethe-Bloch lines for various particle species. The dashed lines 
show the pion and proton exclusion bands. The dotted line corresponds to 
the $+3\sigma$ cut for electrons (see text).}
\label{fig:1}
\end{figure}

The particle identification performance of the TPC  
is essential for the \jpsi\ measurement.
In Fig.~\ref{fig:1} the specific energy loss in the TPC is shown as a function 
of momentum in the region of interest for the present measurement.
A $\pm$3 $\sigma$ inclusion cut for electrons and a $\pm$3.5 $\sigma$ 
($\pm$3 $\sigma$) exclusion cuts for pions (protons) were employed. As seen in Fig.~\ref{fig:1}, with our current identification strategy, the electron identification is performed with an efficiency better than 50\% for momenta below 7-8 GeV/c.

Electron candidates compatible, together with a positron candidate, with being
products of $\gamma$ conversions were removed, in order to reduce the
combinatorial background. It was verified, using a Monte Carlo simulation, 
that this procedure does not affect the \jpsi\ signal.

\begin{figure}[htbp]
\centering{\includegraphics[width=.5\textwidth]{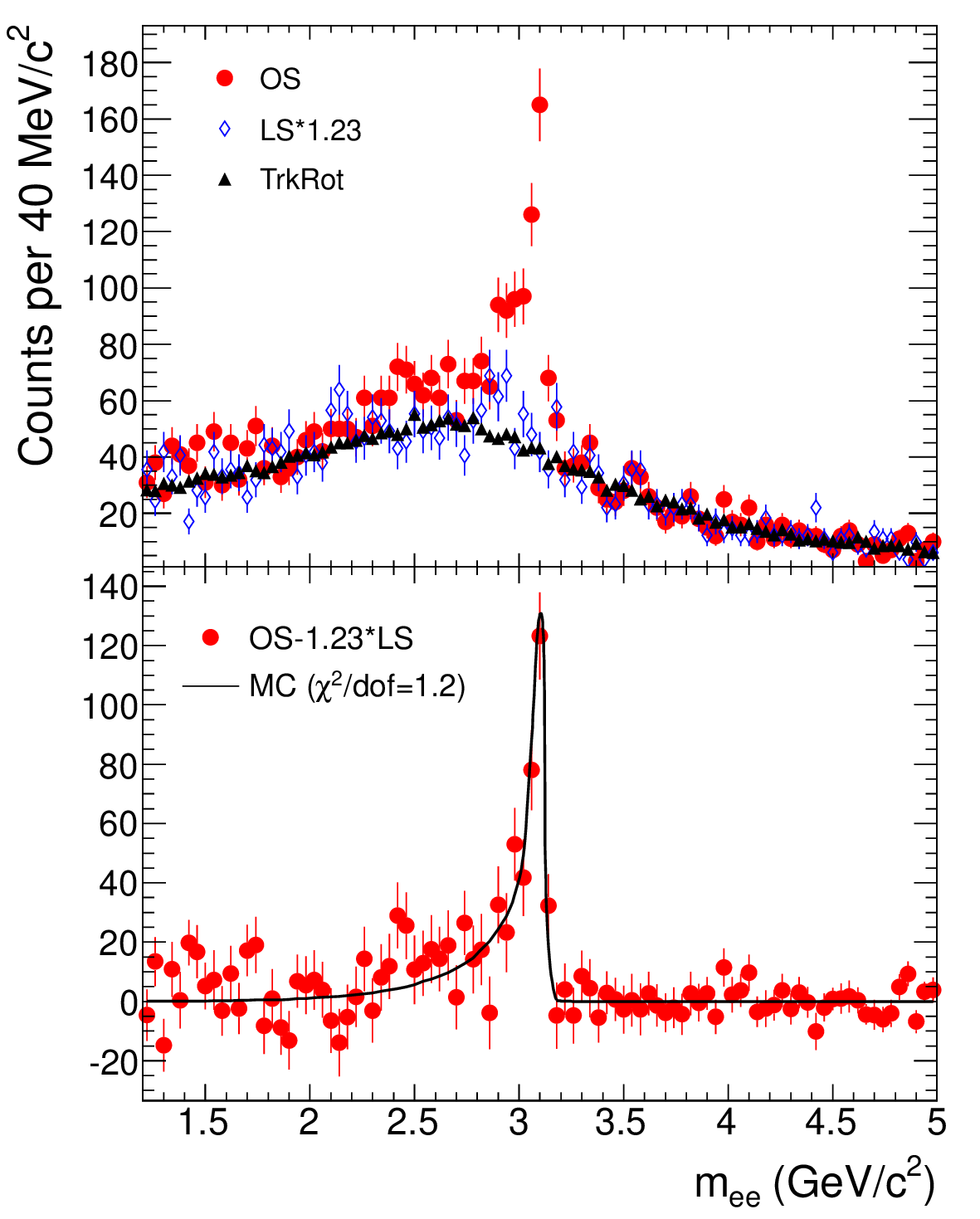}}
\caption{Top panel: invariant mass distributions for opposite-sign (OS) and 
like-sign (LS) electron pairs ($|y|<0.9$, all $p_{\rm T}$), 
as well as for pairs obtained with one track
randomly rotated (TrkRot, see text).  Bottom panel: the difference of the OS 
and LS distributions with the fit to the Monte Carlo (MC) signal 
superimposed.}
\label{fig:2}
\end{figure}

The invariant mass distribution for the opposite-sign (OS) electron pairs is 
shown in Fig.~\ref{fig:2}. In the same figure we also show the background 
contribution, obtained as the sum of the like-sign (LS) pairs, 
${N^{++}+N^{--}}$, scaled to match the integral of the OS distribution in the 
mass interval 3.2$-$5.0 GeV/$c^2$.
The scale factor, 1.23, originates from the presence of correlated background 
(mostly from semileptonic charm decays) in the OS distribution, but is also 
influenced by misidentified electrons and by electrons from conversions.
In the top panel of Fig.~\ref{fig:2} we also show the background estimated
using a track rotation method (TrkRot)~\footnote{The method consists in rotating, 
around the $z$ axis, one of the tracks of the OS pair by a random azimuthal angle. 
More pairs can be obtained by applying the method several times to the same pair.}, 
used later in the estimate of the systematic uncertainties related to signal extraction.
The signal, obtained by subtracting the scaled LS distribution from the OS, is shown in the 
bottom panel of Fig.~\ref{fig:2} in comparison with the signal from Monte 
Carlo (MC) simulations (described below). 
A good agreement between data and MC is observed, both for the bulk of the 
signal and for the bremsstrahlung tail.
Integration of the signal in the mass range 2.92$-$3.16 GeV/$c^2$ yields
$N_{\rm J/\psi}$~=~352~$\pm$~32~(stat.)~$\pm$~28~(syst.) counts (the systematic 
uncertainty on this quantity is described below); the signal to background
ratio is $S/B$~=~1.2~$\pm$~0.1 and the significance is 
$S/\sqrt{S+B}$~=~13.9~$\pm$~0.6.
The tagging and corresponding rejection of $\gamma$ conversions is found to improve $S/B$ by $\sim$30\%.
The MC simulations~\footnote{In the MC simulation the decay of the J/$\psi$ particles 
was handled by the EvtGen package~\cite{evtgen} and the final state radiation was described using PHOTOS~\cite{photos1,photos2}.}
show that 
(66.8~$\pm$~1.9)\% 
of the signal is within the 
integration range. 
The main contribution to the uncertainty on this quantity 
was obtained by  
analyzing MC samples where the detector material budget was varied by 
$\pm$6\%~\cite{Koc11} with respect to the nominal value, and by
varying the track-related cuts ($p_{\rm T}$ and required number of TPC 
clusters) around their nominal values. 
A smaller contribution (1~\%, in terms of the relative error) is attributed to the
small discrepancies between the invariant mass distribution as provided by QED at the
next to leading order~\cite{QED} and by the event generator (EvtGen + PHOTOS); the latter contribution
remains even after taking into account the detector resolution.  
A fit to the invariant mass distribution after background subtraction with a 
Crystal Ball function~\cite{Gai82} gives a mass resolution of 
28.3$\pm$1.8 MeV/$c^2$. 

For the dimuon channel, the total data sample available for physics
analysis amounts to 1.9$\cdot$10$^8$ MB events, of which 
1.0$\cdot$10$^7$ satisfy the $\mu$-MB condition.

An accurate alignment of the tracking chambers of the muon
spectrometer is an essential pre-requisite to identify resonances in
the $\mu^+\mu^-$ invariant mass spectrum. This was carried out
using a modified version of the MILLEPEDE package~\cite{Blo02,Ape09},
starting from a sample of 3$\cdot$10$^5$ tracks, taken with no
magnetic field in the dipole and in the solenoid. The resulting
alignment precision is $\sim$750 $\mu$m in the bending and 
non-bending directions.

Track reconstruction is based on a Kalman filter
algorithm~\cite{Ape09,Cha03}. 
The procedure starts from the most downstream
tracking stations (4 and 5), which are less subject to the background
due to soft particles that escape the front absorber.
Straight line segments are formed by joining clusters on the two planes
of each station and a first estimate of the track parameters
(position, slope and inverse bending momentum) and 
corresponding errors is made. The momentum is first estimated
assuming that the track originates from the vertex and is bent by a
constant magnetic field in the dipole. In a second step, track
candidates on station 4 are extrapolated to station 5 (or vice versa)
and paired with at least one cluster on the basis of a $\chi^2$
cut. If several clusters are found, the track is duplicated to
consider all the possible combinations. After this association the track
parameters and errors are recalculated using the Kalman filter. 

The same procedure is repeated iteratively for the upstream stations, 
rejecting, at each step, the candidates for which no cluster 
is found or those whose parameters indicate that they will exit 
the geometrical acceptance of the spectrometer in the next steps. 
At the end of the procedure,
additional algorithms are applied to improve the track quality by
adding/removing clusters based on a $\chi^2$ cut, and removing fake
tracks sharing clusters with others. Finally, the remaining tracks are
extrapolated to the primary vertex position as given by the SPD~\cite{Aam10}, 
and their parameters are recomputed taking into account the energy loss
and multiple Coulomb scattering in the absorber. With the alignment 
precision obtained for the analyzed data sample the relative momentum
resolution of the reconstructed tracks ranges between 2\% at 10 GeV/$c$ 
and 10\% at 100 GeV/$c$.

After reconstruction, 4.1$\cdot$10$^5$ events having at least
two muon candidates are found, out of which only 6\% have three or more
muons. Various selection cuts are then applied to this data sample.
First, events are required to have at least one interaction vertex
reconstructed by the SPD. This cut rejects 0.5\% of the statistics.
Then, it is required that at least one of the two muon candidates matches
the corresponding hits in the trigger chambers. In this way
hadrons produced in the absorber, which are stopped by the iron wall
positioned upstream of the trigger chambers, are efficiently
rejected. This cut rejects $\sim$24\% of the muon pairs, and its effect is
important only for $m_{\mu\mu}<$1 GeV/$c^2$. In fact, since in 99\% of
the cases at least one of the two \jpsi\ decay muons has a transverse
momentum larger than the trigger $p_{\rm T}$ threshold, the signal
loss induced by this cut is negligible. Requiring both candidate muon
tracks to be matched with the corresponding ``trigger tracks'' would
increase the purity of the muon sample, but it was checked that this
cut would lead to a loss of $\sim$ 20\% of the \jpsi\ events without
decisively increasing the signal to background ratio at the \jpsi\
invariant mass.  Furthermore, the cut $R_{\rm {abs}}>17.5$ cm, where
$R_{\rm {abs}}$ is the radial coordinate of the track at the end of the front
absorber, was applied. In this way, muons emitted at
small angles, that have crossed a significant fraction of the thick beam 
shield, can be rejected. 
Finally, to remove
events very close to the edge of the spectrometer acceptance, the cut
$2.5<y<4$ on the pair rapidity was applied. These quality
cuts reject 10.3\% of the muon pairs.

\begin{figure}[htbp]
\centering
\resizebox{0.6\textwidth}{!}
{\includegraphics*[bb=0 0 565 544]{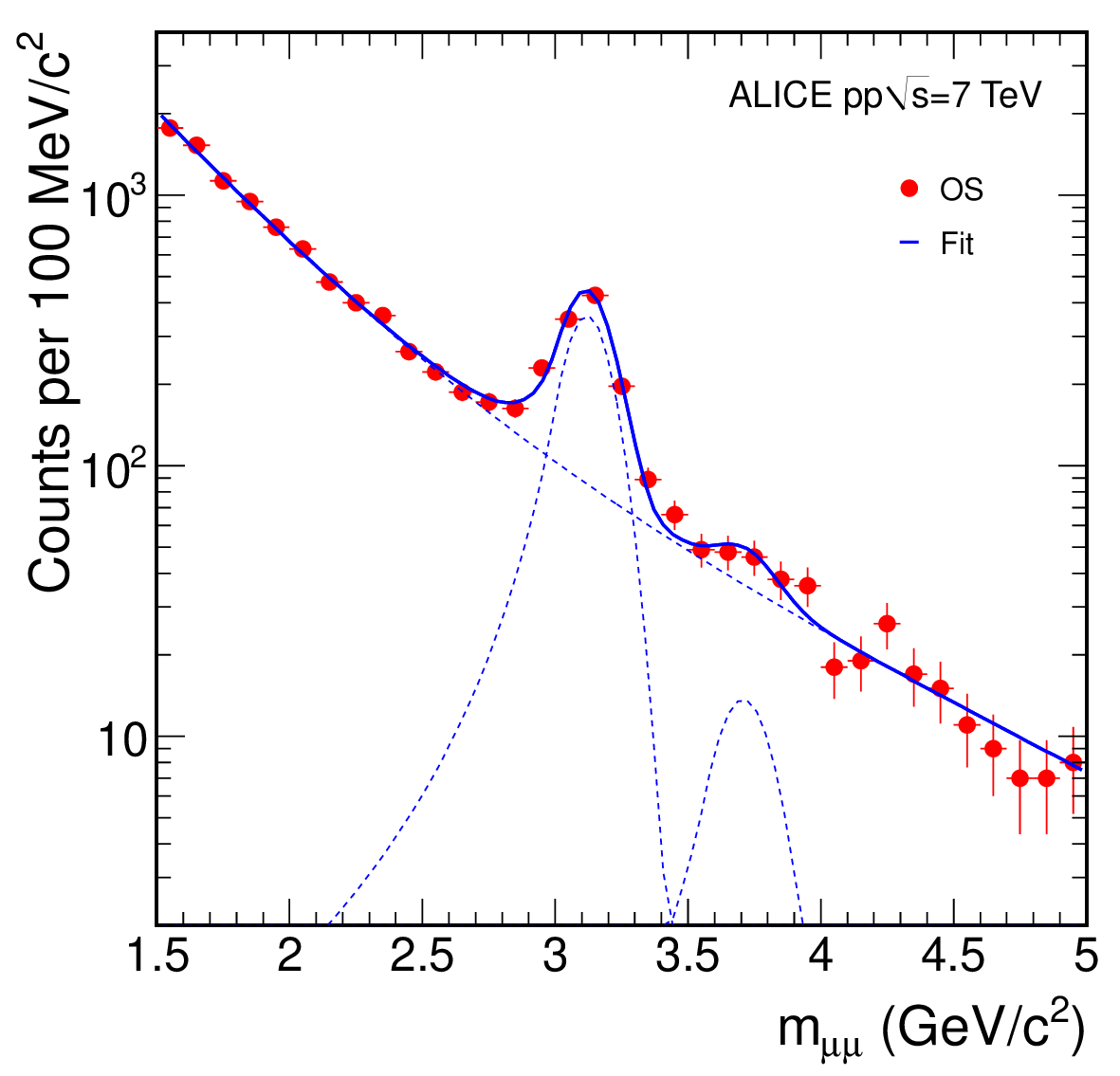}}
\caption{Invariant mass distribution for opposite-sign muon pairs ($2.5<y<4$, all $p_{\rm T}$), 
in the mass region $1.5<m_{\mu\mu}<5$ GeV/$c^2$ with the result of
the fit. The plot refers to the sub-period with the largest statistics 
($N_{\rm J/\psi}=957\pm56$, corresponding to $L_{\rm int}$~=~7.9~nb$^{-1}$).
The fitted J/$\psi$ and $\psi(2S)$ contributions, as well as the 
background, are also shown.}
\label{fig:3}
\end{figure}

After selection, the dimuon sample consists of 1.75$\cdot$10$^5$  
OS muon pairs. 
In Fig.~\ref{fig:3} we present the OS
invariant mass spectrum for the mass region $1.5<m_{\mu\mu}<5$
GeV/$c^2$, corresponding to the sub-period having the
largest statistics. A peak corresponding to the 
J/$\psi\rightarrow\mu^+\mu^-$
decay is clearly visible in the spectrum, on top of a large 
continuum.  A weaker signal, corresponding to the $\psi(2S)$ decay, is
also visible, in spite of the poor signal to background ratio.

The number of signal events $N_{\rm J/\psi}$ was extracted by
fitting the mass range $1.5<m_{\mu\mu}<5$ GeV/$c^2$. The
\jpsi\ and $\psi(2S)$ line shapes are described with Crystal
Ball functions~\cite{Gai82}, while the underlying continuum was
parameterized using the sum of two exponentials.  The functions
representing the resonances were obtained by fitting the expected
mass distribution of a pure MC signal sample.  Such a sample was
obtained by generating, for each sub-period, 
\jpsi\ and $\psi(2S)$ events with realistic differential distributions 
(see below for details). In order to account for small uncertainties in the
MC description of the set-up, the position of the \jpsi\ mass
pole, as well as the width of the Crystal Ball function, were
kept as free parameters in the invariant mass fit. 
Due to the small
statistics, the $\psi(2S)$ parameters were tied to those of the
J/$\psi$, imposing the mass difference between the two states to
be equal to the one given by the Particle Data Group
(PDG)~\cite{Nak10}, and the ratio of the resonance widths to be equal
to the one obtained in the MC.

This choice of parameters leads to a satisfactory fit of the invariant
mass spectrum ($\chi^2/ndf$=1.14), as shown in Fig.~\ref{fig:3}. 
The Crystal Ball function describing the J/$\psi$ is peaked at 
$m_{\rm J/\psi}$~=~3.118~$\pm$~0.005~GeV/$c^2$.  Such a value is 
larger than the one quoted by the PDG group by only 0.6\%, showing that the
accuracy of the magnetic field mapping and of the energy loss
correction is reasonably under control.  The measured width of the
Crystal Ball function is $\sigma_{\rm J/\psi}$~=~94~$\pm$~8~MeV/$c^2$,
in agreement within less than 2\% with the MC, and its FWHM is 221 MeV/$c^2$. 

The same fitting procedure, applied to the other sub-periods, gives consistent
results in terms of both $m_{\rm J/\psi}$ (within 0.2\%) and 
$\sigma_{\rm J/\psi}$ (within 4\%). The signal to background ratio, 
in the mass range $2.9<m_{\mu\mu}<3.3$ GeV/$c^2$, varies between
2.3 and 2.9 in the various sub-periods.
The total number of \jpsi\ signal events, obtained by integrating the 
Crystall Ball function over the full mass range, is
$N_{\rm J/\psi}=1924\pm77 {\rm(stat.)} \pm 144 {\rm(syst.)}$. The
determination of the systematic uncertainty on $N_{\rm J/\psi}$ is described
later in Section~\ref{sec:syst}.

\section{Acceptance and efficiency corrections, luminosity normalization}
\label{sec:acc}

In order to extract the \jpsi\ yield, the number of signal events must 
be corrected, with a MC procedure, for the acceptance of the apparatus 
and for reconstruction and triggering
efficiencies. This procedure is based on the generation of
a large sample of signal events, with a $p_{\rm T}$ distribution
extrapolated from CDF measurements~\cite{Aco05} and a $y$ distribution
parameterized from Color Evaporation Model (CEM)
calculations~\cite{Sto06}. To avoid the
loss of events due to smearing effects at the edge of the angular
acceptance, the generation was performed over
$y$ ranges wider than those covered by the two detector systems.
It was also assumed that \jpsi\ production is
unpolarized. The acceptance factors are obtained with respect to the \jpsi\ rapidity ranges $|y|<0.9$ and $2.5<y<4.0$ for the central barrel and muon detectors, respectively.

For the central barrel detectors, the acceptance times efficiency value 
($A \times \epsilon$) is 
8.5\% 
and is the product of four
contributions: 
i) a kinematic factor, 
namely the requirement of having both e$^+$ and e$^-$  within the acceptance 
($|\eta^{{\rm e}^+,{\rm e}^-}|<0.9$), satisfying a transverse momentum cut 
$p_{\rm T}^{{\rm e}^+,{\rm e}^-}>1$ GeV/$c$.
This factor amounts to 
31.3\%;
ii) the reconstruction efficiency for the e$^+$e$^-$ pair, which is 50.3\%;
iii) the identification efficiency, which is 81.0\%; iv) the fraction of the 
signal within the mass range 2.92-3.16 GeV/$c^2$, which is 
66.8\%.
  
For the muon spectrometer, the tracking efficiency is calculated 
from MC simulations, including a realistic map of dead channels and the residual 
misalignment of the detection elements. This efficiency, obtained from a sample of generated tracks which match the hit pattern on the various chambers
required by the reconstruction algorithm, is $(97.1 \pm 0.8)$\%.

The efficiencies of the muon trigger chambers are 
obtained from the analysis of the ``trigger tracks'' collected in the measured
data sample. The ``trigger tracks'', as explained in Section 2, are defined by the presence of a hit in at least 3 (out of 4) trigger planes. The efficiency for a chamber belonging to a certain trigger plane is calculated starting from a sample of ``trigger tracks'' where the corresponding chambers on the other 3 planes have recorded a hit, and then looking for the presence or absence of a hit in the chamber under study. Such a requirement on the sample does not introduce a bias 
in the efficiency calculation because the response of the detector planes are independent. 
Typical efficiency values are around 96\%,
with 90\% of the detector surface having an efficiency larger than
91\%. The obtained efficiencies are then plugged in the simulations in order to
provide a realistic description of the detector. 
The time variation of the  
tracking and trigger detector efficiencies was accounted for in the 
simulation. Internally to each
sub-period, the response of the tracking chamber channels is further
weighted run by run.

For each sub-period $i$, the ratio between the total number of
reconstructed events, satisfying the analysis cuts, and the generated
events in the range 2.5$< y <$4 gives the product
$A\times\epsilon_{\rm i}$  for the
\jpsi. The differences between the $A\times\epsilon_{\rm i}$ for the various
sub-periods do not exceed 8\%, and their average value is 
$\langle A\times\epsilon\rangle = 32.9$\%.
It is worth noting that $A \times \epsilon$ exhibits, for both the central
barrel detectors and the muon spectrometer, a rather small 
variation as a function of the \jpsi\ $p_{\rm T}$, down to 
zero $p_{\rm T}$.

To get the production cross section value, the ratio 
$N^{\rm {cor}}_{\rm J/\psi}=N_{\rm J/\psi}/\langle A\times\epsilon\rangle$ must be normalized 
to the integrated luminosity, or to the measured cross section for a
chosen reference process. For this analysis, the adopted reference 
is the occurrence of the MB condition itself.
One has simply
\begin{equation}
\sigma_{\rm J/\psi} =  \frac{N^{\rm {cor}}_{\rm J/\psi}}
{BR(\rm J/\psi\rightarrow \ell^+\ell^-)}\times 
\frac{\sigma_{{\rm MB}}}{N_{\rm{MB}}} 
\label{eq:1} 
\end{equation} 
where 
$\rm BR(\rm J/\psi\rightarrow \ell^+\ell^-)$=$(5.94\pm 0.06)$\%~\cite{Nak10}, 
$N_{\rm {MB}}$ is the number of minimum bias collisions and $\sigma_{\rm {MB}}$ 
is the measured cross section for such events. $N_{\rm {MB}}$ was corrected, 
run by run, for the probability of having multiple interactions in a single 
bunch crossing.

In the muon channel, the \jpsi\ signal was collected using the $\mu$-MB trigger
condition. Therefore, Eq.~\ref{eq:1} has to include a multiplicative factor 
$R$ that links the occurrence of a reference process in the $\mu$-MB and MB 
event samples. We have chosen as a reference process the yield $N_{\mu}$ of 
single muons, detected in the region $-4 <\eta< -2.5$, and with $p_{\rm T}>$1 
GeV/$c$. The $R$~factor is then defined as the ratio 
$R=N^{\rm {MB}}_{\mu}/N^{\mu-{\rm {MB}}}_{\mu}$ of the single 
muon yields for the two event samples. 
The numerical values of the $R$~factor strongly depend
on the relative bandwidth assigned by the data acquisition to the two trigger 
samples in each sub-period, and vary between 0.10 and 0.42. However, the choice 
of the $p_{\rm T}$ cut has no significant influence on these values 
($<$1\% for cut values between 0 and 3 GeV/$c$).
This is due to the fact that both the $\mu$-MB and MB muon samples are subject 
to the same set of cuts, including the requirement of matching of the 
reconstructed track with the corresponding trigger track.

The $\sigma_{\rm {MB}}$ value is 62.3 mb, and is affected by a 4\% systematic uncertainty. 
It was obtained relative to the cross section $\sigma_{\rm{V0AND}}$~\cite{Aam12}, 
measured in a van der Meer scan~\cite{vdM68}, of the coincidence V0AND
between signals in the two VZERO detectors. 
The relative factor $\sigma_{\rm{V0AND}}/\sigma_{\rm{MB}}$ was obtained as the 
fraction of MB events where the L0 trigger input corresponding to the V0AND 
condition has fired. Its value is 0.87, and is stable within 
0.5\% over the analyzed data sample.
The integrated luminosity $L_{\rm{int}}=N_{\rm {MB}}/\sigma_{\rm {MB}}$ is 5.6 
nb$^{-1}$ for the dielectron sample.
For the dimuon sample 
$L_{\rm{int}}=(N_{\rm {MB}}/\sigma_{\rm {MB}})/R=15.6$ nb$^{-1}$.

\section{Systematic uncertainties}
\label{sec:syst}

The systematic uncertainty on the inclusive \jpsi\ cross section measurement
was obtained considering the following sources:

\begin{itemize}

\item 
  The uncertainty on the signal extraction procedure, for the electron channel, was 
  estimated using the track rotation method 
  as an alternative background calculation procedure, see Fig.~\ref{fig:2}, 
  and by fitting the invariant mass distributions with a convolution of a polynomial 
  and a Crystal Ball function (with parameters constrained via Monte Carlo). We have 
  also varied the invariant mass ranges for the signal extraction and for background
  normalization. The value obtained is 8\%.  
  Including the contribution from the uncertainty on the material budget 
  and the accuracy of the event generator in describing the radiative decay of the \jpsi\ (J/$\psi\rightarrow$e$^+$e$^-\gamma$)  
  leads 
  to a value of 8.5\%.  
  For the muon channel, various tests were performed. In particular,
  we tried to release in the fit the values of the parameters 
  governing the asymmetric left tail of the Crystal Ball line shape,
  which were fixed to their MC values in the default fitting
  procedure.
  Alternative functions
  for the description of the signal and background shapes were also
  used. In particular, for the \jpsi\ a variable-width Gaussian function, 
  adopted in the past by the NA50 and NA60 Collaborations~\cite{Sha01}, was 
  used. 
  For the background, a different 
  shape was tested, based on a Gaussian having a width continuously increasing
  with the mass. The estimated overall systematic uncertainty on the
  signal extraction is 7.5\%.

\item The acceptance calculation depends on the $y$ and $p_{\rm T}$
  input distributions.  
  For the electrons, the uncertainty is mainly determined by the choice of the
  $p_{\rm T}$ spectrum. By varying the $\langle p_{\rm T} \rangle$ of the input 
  distribution within a factor 2, a 1.5\% variation in the acceptance was obtained.
  Such a small value is indeed a consequence of the weak $p_{\rm T}$ dependence 
  of the acceptance for the bulk of the spectrum. For the muons, both $y$ and 
  $p_{\rm T}$ were varied, using as alternative distributions those expected 
  for $pp$ collisions at $\sqrt{s}$=4 and 10 TeV~\cite{Gri08}. As a further    
  test, the measured J/$\psi$ differential distributions obtained from the 
  analysis described later in Section 6, were used as an input in the 
  calculation of the acceptance. In this way, a 5\% systematic uncertainty on 
  this quantity was determined. The larger systematic uncertainty for the muon
  channel is due to the larger influence, for this channel, of the choice 
  of the shape of the rapidity distribution.
  
\item The uncertainty on the muon trigger efficiency calculation was
  estimated comparing $N^{\rm{cor}}_{\rm J/\psi}$ for the sample where only
  one of the two decay muons is required to match the trigger
  condition, with the same quantity for the sample where both muons
  satisfy that condition. The 4\% discrepancy between the two
  quantities is taken as the systematic uncertainty on the evaluation of
  the trigger efficiency.

\item The uncertainty on the reconstruction efficiency,
  for the central barrel analysis, is due to the track quality
  (4\%) and particle identification (10\%) cuts and originates from residual
  mismatches between data and MC simulations.
  
  For the muon analysis, the systematic uncertainty can be estimated by 
  comparing determinations of the tracking efficiency based on real data and 
  on a MC approach.
  In the first case, the tracking efficiency can be evaluated starting from the 
  determination of the efficiency per chamber, computed using the redundancy 
  of the tracking information in each station. The values thus obtained are in 
  the range from 91.8 to 99.8\%. The tracking efficiency evaluated starting 
  from these chamber efficiencies is $(98.8 \pm 0.8)$\%. The very same 
  procedure, in a MC approach, gives a $(99.8^{+0.2}_{-0.8})$\% tracking 
  efficiency. These two quantities differ by 1\%, which
  is taken as an estimate of the systematic uncertainty. 
  However, this method is not able to detect losses of efficiency due to the 
  presence of correlated dead-areas in the same region of two chambers 
  belonging to the same station. Such correlated dead-areas
  were singled out by studying, on data, the cluster maps of each station, and 
  the corresponding loss of efficiency was estimated to be $(2.8 \pm 0.4)$\%. 
  Taking into account this effect, the resulting tracking efficiency is in 
  good agreement with the value previously quoted (see Section~\ref{sec:acc}) 
  from realistic MC simulations, $(97.1 \pm 0.8)$\%. Nevertheless, a  
  1\% additional systematic error (30\% of the efficiency loss discussed above) 
  was assumed, to take into account possible small-area correlations that could 
  be missed in the present approach.
  Combining this error with the one previously mentioned, the overall 
  systematic uncertainty on the muon tracking efficiency is
  1.5\%, which gives 3\% for muon pair detection. 
   
\item The error on the luminosity measurement is dominated by the 4\% 
  systematic uncertainty on the determination of $\sigma_{\rm{V0AND}}$, which 
  is due to the uncertainties on the beam intensities~\cite{And11} and on the 
  analysis procedure related to the van der Meer scan of the V0AND signal. 
  Other effects, such as the oscillation in the ratio
  between the MB and V0AND counts, contribute to less than 1\%. 
  The cross section $\sigma_{\mu-{\rm MB}}$, relative to the occurence of the 
  $\mu$-MB trigger, was also measured in a van der Meer 
  scan~\cite{Aam12,vdM68} and was used as an alternative reference for the 
  luminosity determination in the muon analysis. 
  Using $\sigma_{\mu-{\rm MB}}$ as a reference cross section, 
  a 4\% difference has been found
  with respect to the integrated luminosity based on $\sigma_{\rm MB}$. For 
  safety, this 4\% has been quadratically added to the 
  luminosity systematic error in the muon analysis.
  In addition, for the muons, the calculation of the integrated luminosity, as described 
  above, is also connected with the estimate of the $R$~factor. This quantity was 
  evaluated in an alternative way, using the information from the trigger 
  scalers and taking into account the dead-time of the triggers.  
  By comparing the two results, a 3\% systematic uncertainty on the $R$~factor 
  was estimated.

\item The branching ratio of the \jpsi\ decay to lepton pairs is 
  known with a 1\% accuracy.

\item The acceptance values significantly depend on the degree of
  polarization assumed in the \jpsi\ distributions.  They were
  calculated in the two cases of fully
  transverse ($\lambda=1$) or longitudinal ($\lambda=-1$)
  polarization\footnote{The polar angle distribution of the \jpsi\ decay 
  leptons is given by $dN/d\cos\theta=1+\lambda\cos^2\theta$}, 
  in the Collins-Soper (CS) and helicity (HE) reference frames. 

\end{itemize}

\begin{table}
\caption{\label{tab:1} Systematic uncertainties (in percent) on the quantities 
 associated to the integrated \jpsi\ cross section measurement.}
\centering
\begin{tabular}{c|c|c|c|c}
\hline
\hline
Channel & \multicolumn{2}{c} {e$^+$e$^-$} & \multicolumn{2}{|c} { $\mu^+\mu^-$} \\ \hline
Signal extraction & \multicolumn{2}{c} {8.5} & \multicolumn{2}{|c} {7.5} \\ \hline
Acceptance input & \multicolumn{2}{c} {1.5} & \multicolumn{2}{|c} {5} \\ \hline
Trigger efficiency & \multicolumn{2}{c} {0} & \multicolumn{2}{|c} {4} \\ \hline
Reconstruction efficiency & \multicolumn{2}{c} {11} & \multicolumn{2}{|c} {3} \\ \hline
R factor & \multicolumn{2}{c} {$-$} & \multicolumn{2}{|c} {3} \\ \hline
Luminosity & \multicolumn{2}{c} {4} & \multicolumn{2}{|c} {5.5}\\ \hline
B.R. & \multicolumn{4}{c} {1} \\ \hline
Polarization & $\lambda=-1$ & $\lambda=1$ & $\lambda=-1$ & $\lambda=1$ \\ \hline
CS & +19 & --13 & +31 & --15 \\ \hline
HE & +21 & --15 & +22 & --10 \\ \hline
\hline
\end{tabular}
\end{table}

The systematic uncertainties are summarized in Table~\ref{tab:1}.
The systematic uncertainty on the inclusive \jpsi\ cross section is
obtained by quadratically combining the errors from the sources
described above, except polarization, and is 12.1\% for the dimuon channel
and 14.5\% for the dielectron one.
The systematic uncertainty due to the unknown \jpsi\ polarization will be 
quoted separately.

\section{Integrated and differential J/$\psi$ cross sections}

The inclusive \jpsi\ production cross sections in $pp$ 
collisions at $\sqrt{s}$=7 TeV are: \\
$\sigma_{\rm J/\psi}(|y|<0.9)$~=~12.4~$\pm$~1.1~(stat.)~$\pm$~1.8~(syst.)
$+$~1.8~($\lambda_{\rm{HE}}=1$)~$-$~2.7~($\lambda_{\rm{HE}}=-1$)~$\mu$b and \\
$\sigma_{\rm J/\psi}(2.5<y<4)$~=~6.31~$\pm$~0.25~(stat.)~$\pm$~0.76~(syst.)
$+$ 0.95~($\lambda_{\rm{CS}}=1$)~$-$~1.96~($\lambda_{\rm{CS}}=-1$)~$\mu$b.

\begin{figure}[htbp]
\centering\resizebox{0.7\textwidth}{!}
{\includegraphics*[bb=0 0 565 662]{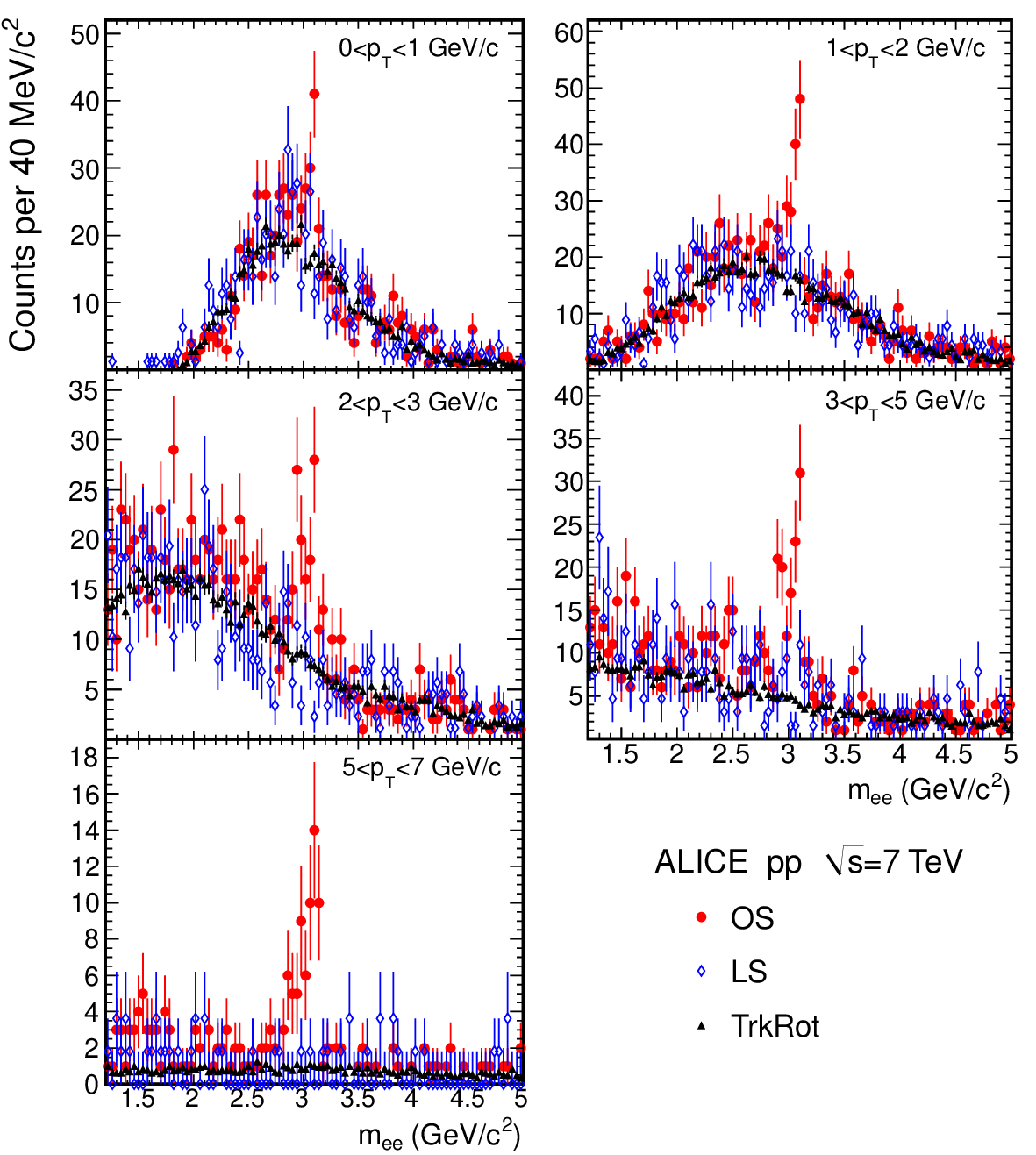}}
\caption{Invariant mass spectra for OS electron pairs ($|y|<0.9$), in bins of $p_{\rm T}$.
The background calculated using LS and TrkRot approaches are also shown.}
\label{fig:elept}
\end{figure}

\begin{figure}[htbp]
\centering\resizebox{1.0\textwidth}{!}
{\includegraphics*[bb=0 0 565 325]{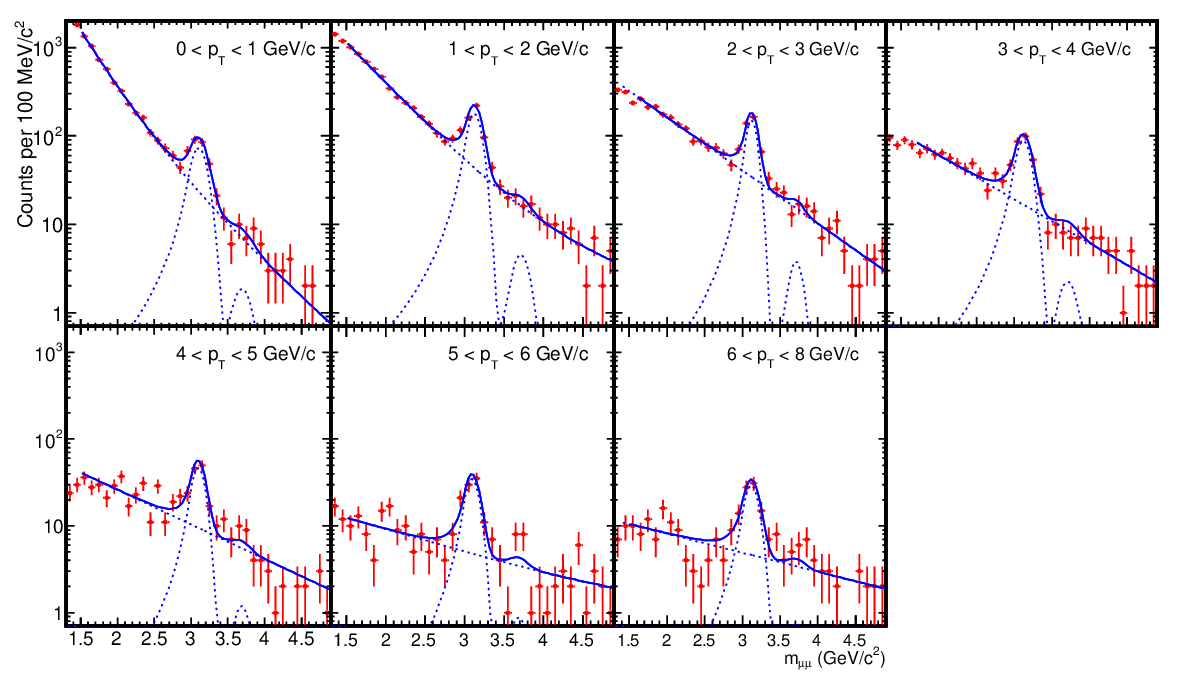}}
\caption{Invariant mass spectra for OS muon pairs ($2.5<y<4$), in bins of $p_{\rm T}$.
The results of the fits are also shown.}
\label{fig:mupt}
\end{figure}

\begin{figure}[ht]
\centering\resizebox{0.6\textwidth}{!}
{\includegraphics*[bb=0 0 565 546]{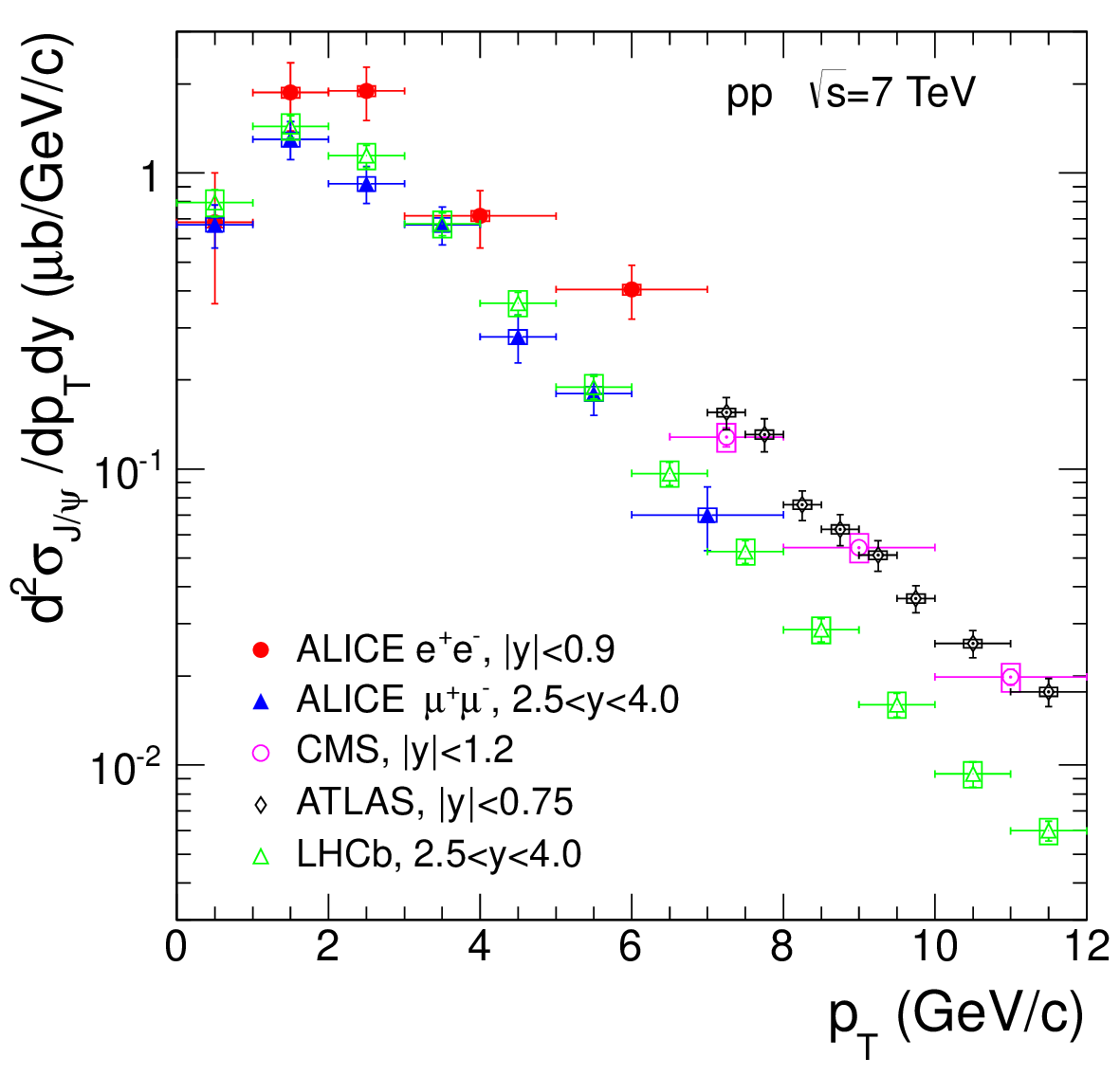}}
\caption{$\mathrm{d}^2\sigma_{\rm J/\psi}/\mathrm{d}p_{\rm T}\mathrm{d}y$ for 
the midrapidity range and for the forward rapidity data, compared with results 
from the other LHC experiments~\cite{Kha10,Aai11,Aad11}, obtained in similar 
rapidity ranges. The error bars represent the quadratic sum of the statistical 
and systematic errors, while the systematic 
uncertainties on luminosity are shown as boxes. 
The symbols are plotted at the center of each bin.}
\label{fig:4}
\end{figure}

The systematic uncertainties related to the unknown polarization are quoted for
the reference frame where they are larger. 

\begin{figure}[htbp]
\centering\resizebox{0.6\textwidth}{!}
{\includegraphics*[bb=0 0 565 546]{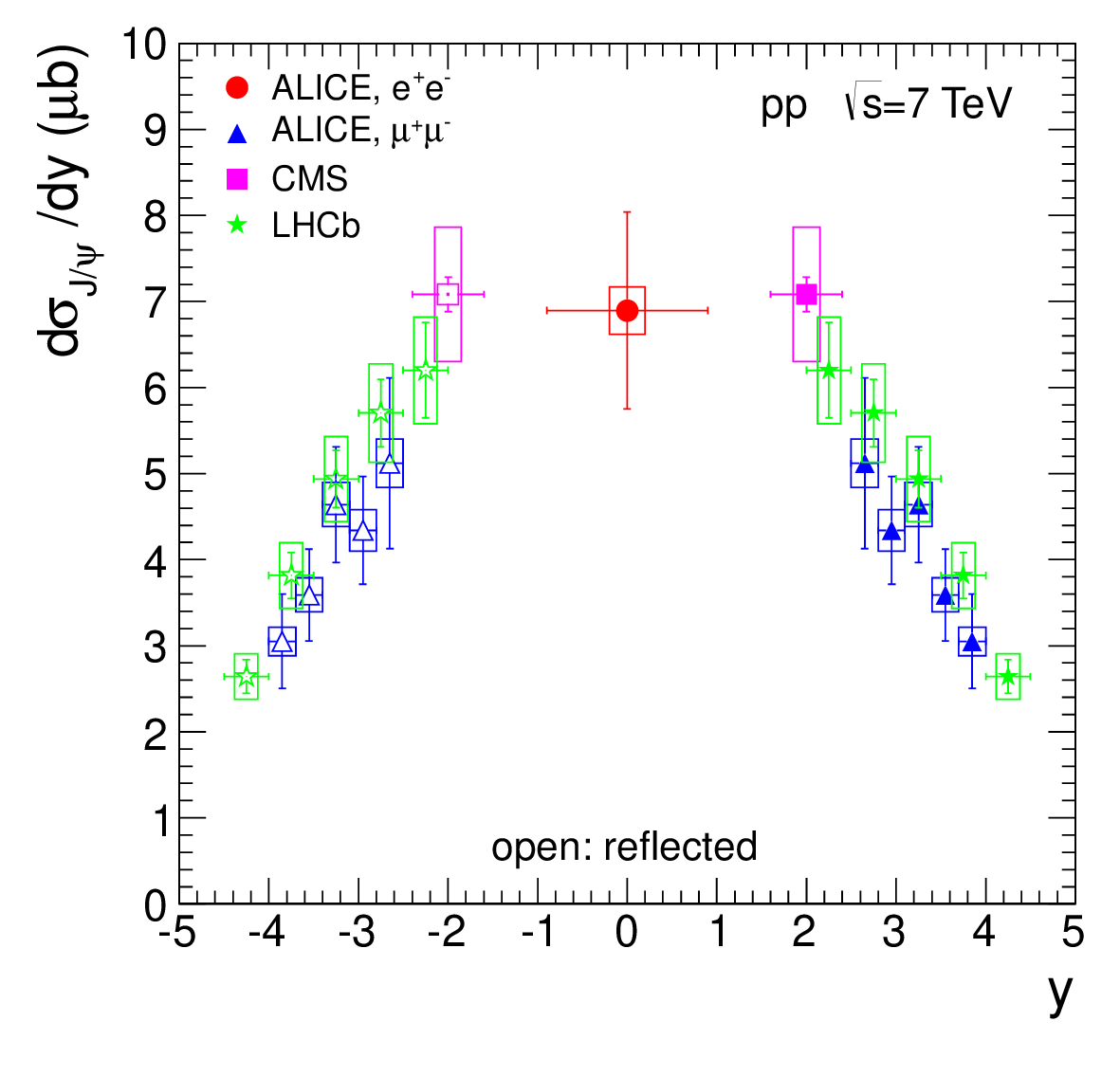}}
\caption{$\mathrm{d}\sigma_{\rm J/\psi}/\mathrm{d}y$, compared with results 
from the other LHC experiments~\cite{Kha10,Aai11,Aad11}.
The error bars represent the quadratic sum of the statistical and systematic 
errors, while the systematic uncertainties on luminosity are shown as boxes.
The symbols are plotted at the center of each bin.}
\label{fig:5}
\end{figure}

In the dielectron channel, the 
$\mathrm{d}\sigma_{\rm J/\psi}/\mathrm{d}p_{\rm T}$ differential cross
section was measured in five $p_{\rm T}$ bins, between 0 and 7 GeV/$c$. In each 
bin, the signal was extracted with the same approach used for the integrated 
invariant mass spectrum. 
In Fig.~\ref{fig:elept} the OS invariant mass spectra are shown, together with the
LS and the TrkRot backgrounds.
The corrections for acceptance and reconstruction 
efficiency and the systematic errors are given in Table~\ref{tab:2}.
Some of the contributions to the systematic uncertainty do not depend on 
$p_{\rm T}$, thus affecting only the overall normalization, and they are 
separately quoted in Table~\ref{tab:2}.
The contributions which depend on $p_{\rm T}$, even when they are correlated 
bin by bin, were included among the non-correlated systematic errors. 

For the analysis in the dimuon channel, a differential study of \jpsi\ 
production was performed in the two kinematic variables $y$ and $p_{\rm T}$ 
separately. In particular, $\mathrm{d}\sigma_{\rm J/\psi}/\mathrm{d}p_{\rm T}$ 
was studied in seven bins between 0 and 8 GeV/$c$, and 
$\mathrm{d}\sigma_{\rm J/\psi}/\mathrm{d}y$ in five bins
between 2.5 and 4. The event sample used for the determination of the 
differential cross sections is slightly smaller (by about 15\%, 
corresponding to $L_{\rm int}$=13.3 nb$^{-1}$) than 
the one analyzed for the integrated cross section. This is due to the 
fact that the statistics in one of the three sub-periods of the data taking is 
too small to allow a satisfactory fit of the differential invariant mass 
spectra. 

The \jpsi\ signal was extracted, for each $y$
or $p_{\rm T}$ bin, with the same fitting technique used for the
integrated invariant mass spectra. Since the $\psi(2S)$ yield is
rather small and cannot be safely constrained by the data themselves,
 its contribution was fixed in such a way as to have the same
$\psi(2S)/(\rm J/\psi)$ ratio extracted from the integrated spectrum. 
Anyway, the results of the fit, for what concerns $N_{{\rm J}/\psi}$, 
are quite insensitive to the precise level of the $\psi(2S)$ contribution. 
It has been verified, for example, that fixing for each $p_{\rm T}$ bin the ratios 
$\psi(2S)/({\rm J}/\psi)$ to the values measured (in the range $p_{\rm T} > 2 $~GeV/c) 
by CDF~\cite{Aal09}, $N_{{\rm J}/\psi}$ varies by less than 1\%.
In Fig.~\ref{fig:mupt} the OS invariant mass spectra are shown, together with the
result of the fits.

The acceptance times efficiency was calculated
differentially in $y$ and $p_{\rm T}$ and the values are reported in
Table~\ref{tab:2}.  It can be noted that as a function of $p_{\rm T}$, 
the $A\times\epsilon$ coverage of
the muon spectrometer for \jpsi\ production extends down to zero
$p_{\rm T}$, and that the values vary by less than a
factor 1.6 in the analyzed $p_{\rm T}$ range. $A\times\epsilon$ has a
stronger $y$ dependence, but its values are larger than 10\%
everywhere. 

The differential cross sections are then calculated with the same
approach used for the integrated cross section, normalizing
$N^{\rm{cor}}_{\rm J/\psi}(y)$ and $N^{\rm{cor}}_{\rm J/\psi}(p_{\rm T})$ to 
the integrated luminosity.
The differential cross sections are affected by the same 
systematic error sources discussed in the previous section. 
All except the one related to the signal extraction can be
considered as common, or strongly correlated.
Table~\ref{tab:2} gives a summary of the results,
including the various sources of systematic uncertainties (correlated, 
uncorrelated and polarization-related). 

The results are presented in Fig.~\ref{fig:4} and Fig.~\ref{fig:5} for
the $p_{\rm T}$-differential cross section
$\mathrm{d}^{2}\sigma_{\rm  J/\psi}/\mathrm{d}p_{\rm T}\mathrm{d}y$
and $\mathrm{d}\sigma_{\rm J/\psi}/\mathrm{d}y$ ($p_{\rm T}>0$), respectively.
For the rapidity distribution, the values obtained in the forward region were 
also reflected with respect to $y=0$. In both figures, the symbols are plotted
at the center of each bin.  The statistical and systematic errors were added in 
quadrature, apart from the 4 (5.5)\% systematic uncertainties on luminosity for the
dielectron (dimuon) channels, shown as boxes.
The differential cross sections shown in Fig.~\ref{fig:4} and~\ref{fig:5}
assume unpolarized \jpsi\ production.  Systematic uncertainties due to the
unknown \jpsi\ polarization are not shown.
Our results are compared with those by the CMS~\cite{Kha10}, LHCb~\cite{Aai11} 
and ATLAS~\cite{Aad11} Collaborations. 
Also for these data the uncertainty due to luminosity, which is 11\% for CMS, 
10\% for LHCb and 3.4\% for ATLAS, is shown separately (boxes), while the 
error bars contain the statistical and the other sources of systematic errors 
added in quadrature.
Our measurement at central rapidity reaches $p_{\rm T}=0$ and is therefore 
complementary to the data of CMS, available at $|y|<1.2$ for $p_{\rm T}>6.5$ 
GeV/$c$, and ATLAS, which covers the region $|y|<0.75$, $p_{\rm T}>7$ GeV/$c$. 
In order to compare our 
$\mathrm{d}^{2}\sigma_{\rm  J/\psi}/\mathrm{d}p_{\rm T}\mathrm{d}y$ in the 
forward rapidity range with that of LHCb, we added the LHCb data for prompt 
and non-prompt production and integrated in the range $2.5<y<4$ to match our 
measurement. The agreement between the two data sets is good.

In Fig.~\ref{fig:5} our results are compared with the corresponding
values from the CMS and LHCb experiments, for the rapidity bins where the
$p_{\rm T}$ coverage extends down to zero (ATLAS has no coverage down to 
$p_{\rm T}$=0 in any rapidity range). 
For CMS, the value for $1.6<|y|<2.4$ was obtained by integrating the 
published $\mathrm{d}^{2}\sigma_{\rm J/\psi}/\mathrm{d}p_{\rm T}\mathrm{d}y$
data~\cite{Kha10}, while for LHCb the published 
$\mathrm{d}\sigma_{\rm J/\psi}/\mathrm{d}y$ for prompt and non-prompt 
production~\cite{Aai11} were added. 
Our data, together with that of the other LHC experiments, constitute a
comprehensive measurement of inclusive \jpsi\ production cross section as
a function of rapidity.
At the LHC, the inclusive \jpsi\ production cross section at central rapidity
is almost twice larger than at Tevatron ($\sqrt{s}$=1.96 TeV)~\cite{Aco05}
and about ten times larger than at RHIC ($\sqrt{s}$=0.2 TeV)~\cite{Ada07}.
The width (FWHM) of the rapidity distribution derived from our data is
about twice larger than at RHIC \cite{Ada07}.

\begin{sidewaystable}
\caption{\label{tab:2}Summary of the results on the \jpsi\ differential cross 
sections.}
\centering
\begin{tabular}{cccccccc}
$p_{\rm T}$ & $N_{\rm J/\psi}$ & A$\times\epsilon$ & $\mathrm{d}^2\sigma_{\rm J/\psi}/\mathrm{d}p_{\rm T}\mathrm{d} y$ & \multicolumn{4}{c} {Systematic errors } \\
 (GeV/c) &  &  & ($\mu$b/(GeV/c)) &  Correl. & Non-correl. & Polariz., CS & Polariz., HE \\
  &  &  &  & {($\mu$b/(GeV/c))} & {($\mu$b/(GeV/c))} & {($\mu$b/(GeV/c))} & {($\mu$b/(GeV/c))}\\
\hline
& \multicolumn{6}{c} {$|y|<0.9$} & \\  
\hline
 $[0; 1]$ & 50$\pm$17 & 0.122 & 0.68$\pm$0.24 & 0.02 & 0.21 & $+0.16,-0.18$ & $+0.08,-0.12$\\
 $[1; 2]$ & 86$\pm$17 & 0.076 & 1.87$\pm$0.37 & 0.07 & 0.31 & $+0.42,-0.50$ & $+0.28,-0.39$\\
 $[2; 3]$ & 79$\pm$13 & 0.069 & 1.89$\pm$0.31 & 0.08 & 0.23 & $+0.33,-0.43$ & $+0.35,-0.44$\\
 $[3; 5]$ & 75$\pm$13 & 0.086 & 0.72$\pm$0.13 & 0.02 & 0.09 & $+0.06,-0.08$ & $+0.16,-0.13$\\
 $[5; 7]$ & 50$\pm$9  & 0.104 & 0.40$\pm$0.07 & 0.01 & 0.05 & $+0.001,-0.005$ & $+0.06,-0.08$\\
\hline
& \multicolumn{6}{c} {$2.5<y<4$} & \\  
\hline
 $[0; 1]$ & 229$\pm$29 & 0.280 & 0.67$\pm$0.08 & 0.06 & 0.05 &  $+0.14,-0.21$  &  $+0.13,-0.19$  \\
 $[1; 2]$ & 453$\pm$40 & 0.287 & 1.30$\pm$0.11 & 0.12 & 0.10 &  $+0.31,-0.35$  &  $+0.19,-0.29$  \\
 $[2; 3]$ & 324$\pm$26 & 0.289 & 0.92$\pm$0.07 & 0.09 & 0.07 &  $+0.17,-0.26$  &  $+0.09,-0.19$  \\
 $[3; 4]$ & 253$\pm$21 & 0.312 & 0.67$\pm$0.06 & 0.06 & 0.05 &  $+0.12,-0.18$  &  $+0.07,-0.11$  \\
 $[4; 5]$ & 120$\pm$17 & 0.359 & 0.28$\pm$0.04 & 0.03 & 0.02 &  $+0.04,-0.05$  &  $+0.02,-0.03$  \\
 $[5; 6]$ & 86$\pm$12  & 0.392 & 0.18$\pm$0.02 & 0.02 & 0.01 &  $+0.01,-0.03$  &  $+0.01,-0.02$  \\
 $[6; 8]$ & 80$\pm$12  & 0.452 & 0.07$\pm$0.01 & 0.01 & 0.01 &  $+0.007,-0.003$ & $+0.007,-0.008$  \\
\hline
\hline
$y$ &  &  & $\mathrm{d}\sigma_{\rm J/\psi}/\mathrm{d} y$ ($\mu$b) & ($\mu$b) & ($\mu$b) &  ($\mu$b) &  ($\mu$b)\\
\hline
$[-0.9; 0.9]$ & 352$\pm$32 & 0.085 & 6.90$\pm$0.62  & 0.28 & 0.96 & $+0.9, -1.3$  & $+1.0, -1.5$\\
 $[2.5; 2.8]$ & 272$\pm$28 & 0.117 & 5.12$\pm$0.77 & 0.49 & 0.38 & $+1.29, -1.62$ & $+0.94, -1.29$ \\
 $[2.8; 3.1]$ & 326$\pm$30 & 0.383 & 4.34$\pm$0.34 & 0.41 & 0.32 & $+0.97, -1.06$ & $+0.85, -0.99$ \\
 $[3.1; 3.4]$ & 409$\pm$32 & 0.469 & 4.64$\pm$0.35 & 0.44 & 0.35 & $+0.53, -0.92$ & $+0.46, -0.87$ \\
 $[3.4; 3.7]$ & 271$\pm$26 & 0.417 & 3.59$\pm$0.30 & 0.34 & 0.27 & $+0.57, -0.77$ & $+0.22, -0.53$ \\
 $[3.7; 4.0]$ & 172$\pm$23 & 0.215 & 3.05$\pm$0.40 & 0.29 & 0.23 & $+0.67, -1.01$ & $+0.09, -0.46$ \\
\end{tabular}										   
\end{sidewaystable}

We stress that the results described in this Letter refer to inclusive
\jpsi\ production. Therefore the measured yield is a 
superposition of a direct component and of \jpsi\ coming from 
the decay of higher-mass charmonium states, in particular the $\chi_{\rm{c1}}$, 
$\chi_{\rm{c2}}$ and $\psi(2S)$ states. These contributions were measured in lower-energy experiments and were found to be $\sim$25\% ($\chi_{\rm{c1}}+\chi_{\rm{c2}}$) and 
$\sim$8\% ($\psi(2S)$) of the total measured \jpsi\ yield~\cite{Fac08,Ada11}. The $\chi_{\rm{c0}}$ contribution 
is negligible since its B.R. into \jpsi\ is of the order of 1\%.
 In addition to
this ``prompt'' production, decays of beauty hadrons are also
known to give a sizeable contribution (of the order of 10-15\% in the $p_{\rm T}$
range accessed by ALICE~\cite{Aai11}) 
 to the observed \jpsi\ yield.
With future high-statistics data samples, the ALICE experiment will identify, 
at central rapidities, \jpsi\ from b-decays, via the measurement of the 
pseudo-proper decay time distributions~\cite{Bru10}, and will also reconstruct 
the $\chi_{\rm c}\rightarrow \rm J/\psi+\gamma$ decay~\cite{Lad09}. 
At forward rapidity, the contribution from b-decays will be estimated from
the beauty cross section measurement carried out in the semi-leptonic decay channel~\cite{Sto10}.

\section {Conclusions}

The ALICE experiment has measured inclusive \jpsi\ production in the
rapidity ranges $|y|<0.9$ and $2.5<y<4$, through the decays
J/$\psi\rightarrow {\rm e}^+{\rm e}^-$ and J/$\psi\rightarrow \mu^+\mu^-$,
respectively.  The $p_{\rm T}$-integrated cross sections, based on
data samples corresponding to integrated luminosities  
$L_{\rm{int}}=5.6$ nb$^{-1}$ (for the J/$\psi\rightarrow {\rm e}^+{\rm e}^-$ 
channel) and $L_{\rm int}=15.6$ nb$^{-1}$ (for J/$\psi\rightarrow \mu^+\mu^-$) 
are 
$\sigma_{\rm J/\psi}(|y|<0.9)$ = 12.4 $\pm$ 1.1 (stat.) $\pm$ 1.8 (syst.) $+$ 1.8 ($\lambda_{\rm{HE}}=1$) $-$ 2.7 ($\lambda_{\rm{HE}}=-1$) $\mu$b 
and 
$\sigma_{\rm J/\psi}(2.5<y<4)$ = 6.31 $\pm$ 0.25 (stat.) $\pm$ 0.76 (syst.) $+$ 0.95 ($\lambda_{\rm{CS}}=1$) $-$ 1.96 ($\lambda_{\rm{CS}}=-1$) $\mu$b.  
The transverse momentum distribution was
measured at both central and forward rapidity. 
Taking  together the results from the muon and electron channels, the ALICE 
measurement of the inclusive \jpsi\ production cross 
section is particularly relevant in the context of charmonium studies at the 
LHC, for its coverage of both central and forward rapidities and for the 
lowest $p_{\rm T}$ reach at $y=0$.

%
\newenvironment{acknowledgement}{\relax}{\relax}
\begin{acknowledgement}
\section{Acknowledgements}
The ALICE collaboration would like to thank all its engineers and technicians for their invaluable contributions to the construction of the experiment and the CERN accelerator teams for the outstanding performance of the LHC complex.
The ALICE collaboration acknowledges the following funding agencies for their support in building and
running the ALICE detector:
Calouste Gulbenkian Foundation from Lisbon and Swiss Fonds Kidagan, Armenia;
Conselho Nacional de Desenvolvimento Cient\'{\i}fico e Tecnol\'{o}gico (CNPq), Financiadora de Estudos e Projetos (FINEP),
Funda\c{c}\~{a}o de Amparo \`{a} Pesquisa do Estado de S\~{a}o Paulo (FAPESP);
National Natural Science Foundation of China (NSFC), the Chinese Ministry of Education (CMOE)
and the Ministry of Science and Technology of China (MSTC);
Ministry of Education and Youth of the Czech Republic;
Danish Natural Science Research Council, the Carlsberg Foundation and the Danish National Research Foundation;
The European Research Council under the European Community's Seventh Framework Programme;
Helsinki Institute of Physics and the Academy of Finland;
French CNRS-IN2P3, the `Region Pays de Loire', `Region Alsace', `Region Auvergne' and CEA, France;
German BMBF and the Helmholtz Association;
Greek Ministry of Research and Technology;
Hungarian OTKA and National Office for Research and Technology (NKTH);
Department of Atomic Energy and Department of Science and Technology of the Government of India;
Istituto Nazionale di Fisica Nucleare (INFN) of Italy;
MEXT Grant-in-Aid for Specially Promoted Research, Ja\-pan;
Joint Institute for Nuclear Research, Dubna;
 %
National Research Foundation of Korea (NRF);
CONACYT, DGAPA, M\'{e}xico, ALFA-EC and the HELEN Program (High-Energy physics Latin-American--European Network);
Stichting voor Fundamenteel Onderzoek der Materie (FOM) and the Nederlandse Organisatie voor Wetenschappelijk Onderzoek (NWO), Netherlands;
Research Council of Norway (NFR);
Polish Ministry of Science and Higher Education;
National Authority for Scientific Research - NASR (Autoritatea Na\c{t}ional\u{a} pentru Cercetare \c{S}tiin\c{t}ific\u{a} - ANCS);
Federal Agency of Science of the Ministry of Education and Science of Russian Federation, International Science and
Technology Center, Russian Academy of Sciences, Russian Federal Agency of Atomic Energy, Russian Federal Agency for Science and Innovations and CERN-INTAS;
Ministry of Education of Slovakia;
CIEMAT, EELA, Ministerio de Educaci\'{o}n y Ciencia of Spain, Xunta de Galicia (Conseller\'{\i}a de Educaci\'{o}n),
CEA\-DEN, Cubaenerg\'{\i}a, Cuba, and IAEA (International Atomic Energy Agency);
The Ministry of Science and Technology and the National Research Foundation (NRF), South Africa;
Swedish Reseach Council (VR) and Knut $\&$ Alice Wallenberg Foundation (KAW);
Ukraine Ministry of Education and Science;
United Kingdom Science and Technology Facilities Council (STFC);
The United States Department of Energy, the United States National
Science Foundation, the State of Texas, and the State of Ohio.
%
\end{acknowledgement}
\newpage
%
%
\appendix
\section{The ALICE Collaboration}
\label{app:collab}
%
\begingroup
\small
\begin{flushleft}
K.~Aamodt\Irefn{0}\And
A.~Abrahantes~Quintana\Irefn{1}\And
D.~Adamov\'{a}\Irefn{2}\And
A.M.~Adare\Irefn{3}\And
M.M.~Aggarwal\Irefn{4}\And
G.~Aglieri~Rinella\Irefn{5}\And
A.G.~Agocs\Irefn{6}\And
A.~Agostinelli\Irefn{7}\And
S.~Aguilar~Salazar\Irefn{8}\And
Z.~Ahammed\Irefn{9}
% ...
\renewcommand\labelenumi{\textsuperscript{\theenumi}~}
\section*{Affiliation notes}
\renewcommand\theenumi{\roman{enumi}}
\begin{Authlist}
\item \Adef{0}Also at Laboratoire de Physique Corpusculaire (LPC), Clermont Universit\'{e}, Universit\'{e} Blaise Pascal, CNRS--IN2P3, Clermont-Ferrand, France
\item \Adef{1}Also at European Organization for Nuclear Research (CERN), Geneva, Switzerland
\item \Adef{2}Now at Physikalisches Institut, Ruprecht-Karls-Universit\"{a}t Heidelberg, Heidelberg, Germany
\item \Adef{3}Also at  Dipartimento di Fisica dell'Universit\`{a} , Udine, Italy
\item \Adef{4}Also at Dipartimento di Fisica Sperimentale dell'Universit\`{a} and Sezione INFN, Turin, Italy
\item \Adef{5}Also at Benem\'{e}rita Universidad Aut\'{o}noma de Puebla, Puebla, Mexico
\item \Adef{6} Deceased 
\item \Adef{8}Now at European Organization for Nuclear Research (CERN), Geneva, Switzerland
\item \Adef{9}Also at Wayne State University, Detroit, Michigan, United States
\item \Adef{10}Also at Frankfurt Institute for Advanced Studies, Johann Wolfgang Goethe-Universit\"{a}t Frankfurt, Frankfurt, Germany
\item \Adef{11}Also at Fachhochschule K\"{o}ln, K\"{o}ln, Germany
\item \Adef{12}Also at Institute of Experimental Physics, Slovak Academy of Sciences, Ko\v{s}ice, Slovakia
\item \Adef{13}Also at Laboratoire de Physique Subatomique et de Cosmologie (LPSC), Universit\'{e} Joseph Fourier, CNRS-IN2P3, Institut Polytechnique de Grenoble, Grenoble, France
\item \Adef{14}Also at  "Vin\v{c}a" Institute of Nuclear Sciences, Belgrade, Serbia 
\item \Adef{15}Also at Research Division and ExtreMe Matter Institute EMMI, GSI Helmholtzzentrum f\"ur Schwerionenforschung, Darmstadt, Germany
\item \Adef{16}Also at Instituto de Ciencias Nucleares, Universidad Nacional Aut\'{o}noma de M\'{e}xico, Mexico City, Mexico
\item \Adef{17}Also at University of Houston, Houston, Texas, United States
\item \Adef{18}Also at Department of Physics, University of Oslo, Oslo, Norway
\item \Adef{19}Also at Institute of Theoretical Physics, University of Wroclaw, Wroclaw, Poland
\item \Adef{20}Now at Department of Physics, University of Oslo, Oslo, Norway
\item \Adef{21}Also at Eberhard Karls Universit\"{a}t T\"{u}bingen, T\"{u}bingen, Germany
\item \Adef{22}Also at Hua-Zhong Normal University, Wuhan, China
\item \Adef{23}Now at Eberhard Karls Universit\"{a}t T\"{u}bingen, T\"{u}bingen, Germany
\item \Adef{24}Also at Centro Fermi -- Centro Studi e Ricerche e Museo Storico della Fisica ``Enrico Fermi'', Rome, Italy
\end{Authlist}
\section*{Collaboration Institutes}
\renewcommand\theenumi{\arabic{enumi}~}
\begin{Authlist}
\item \Idef{0}Department of Physics and Technology, University of Bergen, Bergen, Norway
\item \Idef{1}Centro de Aplicaciones Tecnol\'{o}gicas y Desarrollo Nuclear (CEADEN), Havana, Cuba
\item \Idef{2}Nuclear Physics Institute, Academy of Sciences of the Czech Republic, \v{R}e\v{z} u Prahy, Czech Republic
\item \Idef{3}Yale University, New Haven, Connecticut, United States
\item \Idef{4}Physics Department, Panjab University, Chandigarh, India
\item \Idef{5}European Organization for Nuclear Research (CERN), Geneva, Switzerland
\item \Idef{6}KFKI Research Institute for Particle and Nuclear Physics, Hungarian Academy of Sciences, Budapest, Hungary
\item \Idef{7}Dipartimento di Fisica dell'Universit\`{a} and Sezione INFN, Bologna, Italy
\item \Idef{8}Instituto de F\'{\i}sica, Universidad Nacional Aut\'{o}noma de M\'{e}xico, Mexico City, Mexico
\item \Idef{9}Variable Energy Cyclotron Centre, Kolkata, India
\item \Idef{10}Department of Physics Aligarh Muslim University, Aligarh, India
\item \Idef{11}Gangneung-Wonju National University, Gangneung, South Korea
\item \Idef{12}Institute for Theoretical and Experimental Physics, Moscow, Russia
\item \Idef{13}Russian Research Centre Kurchatov Institute, Moscow, Russia
\item \Idef{14}Sezione INFN, Turin, Italy
\item \Idef{15}Centro Fermi -- Centro Studi e Ricerche e Museo Storico della Fisica ``Enrico Fermi'', Rome, Italy
\item \Idef{16}Bogolyubov Institute for Theoretical Physics, Kiev, Ukraine
\item \Idef{17}Faculty of Engineering, Bergen University College, Bergen, Norway
\item \Idef{18}Frankfurt Institute for Advanced Studies, Johann Wolfgang Goethe-Universit\"{a}t Frankfurt, Frankfurt, Germany
\item \Idef{19}Dipartimento Interateneo di Fisica `M.~Merlin' and Sezione INFN, Bari, Italy
\item \Idef{20}V.~Fock Institute for Physics, St. Petersburg State University, St. Petersburg, Russia
\item \Idef{21}National Institute for Physics and Nuclear Engineering, Bucharest, Romania
\item \Idef{22}Research Division and ExtreMe Matter Institute EMMI, GSI Helmholtzzentrum f\"ur Schwerionenforschung, Darmstadt, Germany
\item \Idef{23}Department of Physics, Ohio State University, Columbus, Ohio, United States
\item \Idef{24}Rudjer Bo\v{s}kovi\'{c} Institute, Zagreb, Croatia
\item \Idef{25}Sezione INFN, Padova, Italy
\item \Idef{26}Sezione INFN, Bologna, Italy
\item \Idef{27}SUBATECH, Ecole des Mines de Nantes, Universit\'{e} de Nantes, CNRS-IN2P3, Nantes, France
\item \Idef{28}Institut f\"{u}r Kernphysik, Johann Wolfgang Goethe-Universit\"{a}t Frankfurt, Frankfurt, Germany
\item \Idef{29}Laboratoire de Physique Subatomique et de Cosmologie (LPSC), Universit\'{e} Joseph Fourier, CNRS-IN2P3, Institut Polytechnique de Grenoble, Grenoble, France
\item \Idef{30}Departamento de F\'{\i}sica de Part\'{\i}culas and IGFAE, Universidad de Santiago de Compostela, Santiago de Compostela, Spain
\item \Idef{31}Oak Ridge National Laboratory, Oak Ridge, Tennessee, United States
\item \Idef{32}Helsinki Institute of Physics (HIP) and University of Jyv\"{a}skyl\"{a}, Jyv\"{a}skyl\"{a}, Finland
\item \Idef{33}Sezione INFN, Catania, Italy
\item \Idef{34}Commissariat \`{a} l'Energie Atomique, IRFU, Saclay, France
\item \Idef{35}Laboratoire de Physique Corpusculaire (LPC), Clermont Universit\'{e}, Universit\'{e} Blaise Pascal, CNRS--IN2P3, Clermont-Ferrand, France
\item \Idef{36}Institute of Experimental Physics, Slovak Academy of Sciences, Ko\v{s}ice, Slovakia
\item \Idef{37}Dipartimento di Fisica e Astronomia dell'Universit\`{a} and Sezione INFN, Catania, Italy
\item \Idef{38}School of Physics and Astronomy, University of Birmingham, Birmingham, United Kingdom
\item \Idef{39}The Henryk Niewodniczanski Institute of Nuclear Physics, Polish Academy of Sciences, Cracow, Poland
\item \Idef{40}Institut f\"{u}r Kernphysik, Westf\"{a}lische Wilhelms-Universit\"{a}t M\"{u}nster, M\"{u}nster, Germany
\item \Idef{41}Joint Institute for Nuclear Research (JINR), Dubna, Russia
\item \Idef{42}Niels Bohr Institute, University of Copenhagen, Copenhagen, Denmark
\item \Idef{43}Institut Pluridisciplinaire Hubert Curien (IPHC), Universit\'{e} de Strasbourg, CNRS-IN2P3, Strasbourg, France
\item \Idef{44}University of Houston, Houston, Texas, United States
\item \Idef{45}Dipartimento di Fisica Sperimentale dell'Universit\`{a} and Sezione INFN, Turin, Italy
\item \Idef{46}Petersburg Nuclear Physics Institute, Gatchina, Russia
\item \Idef{47}Physics Department, University of Jammu, Jammu, India
\item \Idef{48}Laboratori Nazionali di Frascati, INFN, Frascati, Italy
\item \Idef{49}Dipartimento di Fisica dell'Universit\`{a} and Sezione INFN, Padova, Italy
\item \Idef{50}Faculty of Nuclear Sciences and Physical Engineering, Czech Technical University in Prague, Prague, Czech Republic
\item \Idef{51}Nikhef, National Institute for Subatomic Physics, Amsterdam, Netherlands
\item \Idef{52}Centro de Investigaciones Energ\'{e}ticas Medioambientales y Tecnol\'{o}gicas (CIEMAT), Madrid, Spain
\item \Idef{53}Moscow Engineering Physics Institute, Moscow, Russia
\item \Idef{54}Institute for High Energy Physics, Protvino, Russia
\item \Idef{55}Faculty of Science, P.J.~\v{S}af\'{a}rik University, Ko\v{s}ice, Slovakia
\item \Idef{56}Wayne State University, Detroit, Michigan, United States
\item \Idef{57}Saha Institute of Nuclear Physics, Kolkata, India
\item \Idef{58}Kirchhoff-Institut f\"{u}r Physik, Ruprecht-Karls-Universit\"{a}t Heidelberg, Heidelberg, Germany
\item \Idef{59}Institut de Physique Nucl\'{e}aire d'Orsay (IPNO), Universit\'{e} Paris-Sud, CNRS-IN2P3, Orsay, France
\item \Idef{60}Department of Physics, University of Oslo, Oslo, Norway
\item \Idef{61}Faculty of Mathematics, Physics and Informatics, Comenius University, Bratislava, Slovakia
\item \Idef{62}Russian Federal Nuclear Center (VNIIEF), Sarov, Russia
\item \Idef{63}Physikalisches Institut, Ruprecht-Karls-Universit\"{a}t Heidelberg, Heidelberg, Germany
\item \Idef{64}Physics Department, University of Cape Town, iThemba LABS, Cape Town, South Africa
\item \Idef{65}Hua-Zhong Normal University, Wuhan, China
\item \Idef{66}Secci\'{o}n F\'{\i}sica, Departamento de Ciencias, Pontificia Universidad Cat\'{o}lica del Per\'{u}, Lima, Peru
\item \Idef{67}Dipartimento di Fisica dell'Universit\`{a} and Sezione INFN, Trieste, Italy
\item \Idef{68}Centro de Investigaci\'{o}n y de Estudios Avanzados (CINVESTAV), Mexico City and M\'{e}rida, Mexico
\item \Idef{69}Physics Department, Creighton University, Omaha, Nebraska, United States
\item \Idef{70}Universit\'{e} de Lyon, Universit\'{e} Lyon 1, CNRS/IN2P3, IPN-Lyon, Villeurbanne, France
\item \Idef{71}Universidade Estadual de Campinas (UNICAMP), Campinas, Brazil
\item \Idef{72}Nikhef, National Institute for Subatomic Physics and Institute for Subatomic Physics of Utrecht University, Utrecht, Netherlands
\item \Idef{73}Division of Experimental High Energy Physics, University of Lund, Lund, Sweden
\item \Idef{74}University of Tsukuba, Tsukuba, Japan
\item \Idef{75}Sezione INFN, Cagliari, Italy
\item \Idef{76}Benem\'{e}rita Universidad Aut\'{o}noma de Puebla, Puebla, Mexico
\item \Idef{77}Dipartimento di Scienze e Tecnologie Avanzate dell'Universit\`{a} del Piemonte Orientale and Gruppo Collegato INFN, Alessandria, Italy
\item \Idef{78}Instituto de Ciencias Nucleares, Universidad Nacional Aut\'{o}noma de M\'{e}xico, Mexico City, Mexico
\item \Idef{79}Institute of Space Sciences (ISS), Bucharest, Romania
\item \Idef{80}Institute of Physics, Bhubaneswar, India
\item \Idef{81}Universidade de S\~{a}o Paulo (USP), S\~{a}o Paulo, Brazil
\item \Idef{82}Dipartimento di Fisica `E.R.~Caianiello' dell'Universit\`{a} and Gruppo Collegato INFN, Salerno, Italy
\item \Idef{83}Sezione INFN, Bari, Italy
\item \Idef{84}Dipartimento di Fisica dell'Universit\`{a} and Sezione INFN, Cagliari, Italy
\item \Idef{85}Soltan Institute for Nuclear Studies, Warsaw, Poland
\item \Idef{86}Sezione INFN, Rome, Italy
\item \Idef{87}Institute for Nuclear Research, Academy of Sciences, Moscow, Russia
\item \Idef{88}Sezione INFN, Trieste, Italy
\item \Idef{89}Physics Department, University of Athens, Athens, Greece
\item \Idef{90}Chicago State University, Chicago, United States
\item \Idef{91}Warsaw University of Technology, Warsaw, Poland
\item \Idef{92}Universidad Aut\'{o}noma de Sinaloa, Culiac\'{a}n, Mexico
\item \Idef{93}Technical University of Split FESB, Split, Croatia
\item \Idef{94}Yerevan Physics Institute, Yerevan, Armenia
\item \Idef{95}University of Tokyo, Tokyo, Japan
\item \Idef{96}Department of Physics, Sejong University, Seoul, South Korea
\item \Idef{97}Lawrence Berkeley National Laboratory, Berkeley, California, United States
\item \Idef{98}Indian Institute of Technology, Mumbai, India
\item \Idef{99}Institut f\"{u}r Kernphysik, Technische Universit\"{a}t Darmstadt, Darmstadt, Germany
\item \Idef{100}Yonsei University, Seoul, South Korea
\item \Idef{101}Zentrum f\"{u}r Technologietransfer und Telekommunikation (ZTT), Fachhochschule Worms, Worms, Germany
\item \Idef{102}California Polytechnic State University, San Luis Obispo, California, United States
\item \Idef{103}China Institute of Atomic Energy, Beijing, China
\item \Idef{104}Institute of Physics, Academy of Sciences of the Czech Republic, Prague, Czech Republic
\item \Idef{105}The University of Texas at Austin, Physics Department, Austin, TX, United States
\item \Idef{106}University of Tennessee, Knoxville, Tennessee, United States
\item \Idef{107}Dipartimento di Fisica dell'Universit\`{a} `La Sapienza' and Sezione INFN, Rome, Italy
\item \Idef{108}Hiroshima University, Hiroshima, Japan
\item \Idef{109}Budker Institute for Nuclear Physics, Novosibirsk, Russia
\item \Idef{110}Physics Department, University of Rajasthan, Jaipur, India
\item \Idef{111}Laboratori Nazionali di Legnaro, INFN, Legnaro, Italy
\item \Idef{112}Purdue University, West Lafayette, Indiana, United States
\item \Idef{113}Lawrence Livermore National Laboratory, Livermore, California, United States
\item \Idef{114}Pusan National University, Pusan, South Korea
\item \Idef{115}Centre de Calcul de l'IN2P3, Villeurbanne, France 
\end{Authlist}
\endgroup
%
%
\end{document}